\documentclass[letterpaper]{appolb}
\usepackage{epsfig}
\usepackage{amsmath}
\usepackage{amsfonts}

\newcommand \tie {{\it i.e.}}

\newcommand \kd  {\delta}
\newcommand \ra  {\rightarrow}

\newcommand \x {\cdot}

\newcommand \bvec{\left( \begin{array}{c} }
\newcommand \evec{\end{array} \right)}

\newcommand \bea{\begin{eqnarray} }
\newcommand \eea{\end{eqnarray} }

\newcommand {\be} {\begin{equation}}
\newcommand {\ee} {\end{equation}}

\begin{document}
\title{The study of dense matter through jet modification
\thanks{Presented at ISMD 2007, LBNL, Berkeley, California, USA; Aug. 4-9 2007.}%
}
\author{A. Majumder
\address{Department of Physics, Duke University, Durham North Carolina, 27708, USA}
}
\maketitle
\vspace{-1cm}
\begin{abstract}
The use of jet modification to study the properties of dense matter is reviewed. Different 
sets of jet correlations measurements which may be used to obtain both the space-time 
and momentum space structure of the produced matter are outlined.
\end{abstract}
 \PACS{12.38.Mh, 11.10.Wx, 25.75.Dw}
 \section{Introduction}
It is now established that the collision of heavy-ions at the Relativistic Heavy-Ion collider have led to the 
formation of an entirely new form of matter~\cite{white_papers}. While the underlying degrees of 
freedom prevalent in the hot plasma are, as yet, unknown~\cite{Shuryak:2003ty}, 
various constraints may be imposed 
through a study of its partonic substructure. The foremost tool in this study 
is the modification of the hard jets, 
usually referred to as jet quenching~\cite{quenching}. 
The number of hadrons with transverse momentum $p_T \geq 7$ GeV (which, necessarily originate in the fragmentation of hard jets) 
is reduced by almost a factor of 5 in central $Au$-$Au$ collisions, compared to that 
expected from elementary nucleon nucleon encounters 
enhanced by the number of expected binary collisions~\cite{Adcox:2001jp}. 

Jet modification is a probe with a wide range of complexity in terms of experimental observables. 
By now, measurements on single inclusive observables have been extended to very high $p_T$
($p_T \leq 20$ GeV).  There also exist, a large number of  multi-particle jet-like correlation observables, 
photon-jet, jet-medium and heavy flavor 
observables~\cite{white_papers}. In these proceedings, we attempt a very brief review of the 
underlying theory and some of the new jet correlation observables which may be used to 
understand the underlying space-time and momentum space structure of the produced matter.


\section{Jet Quenching: theory}


Most current calculations of the in-medium modification of light partons may 
be divided into four major schemes, often referred to by the names of the original 
authors.
All schemes utilize a factorized approach, where the 
final cross section to produce a hadron $h$ with high transverse momentum $p_T$ 
and a pseudo-rapidity between $\eta$ and $\eta+d\eta$
may be expressed as an integral over the product of the 
nuclear structure functions [$G_a^A(x_a),G_b^B(x_b) $], 
to produce partons with momentum fractions $x_a,x_b$, a 
hard partonic cross section to produce a hard parton with a 
transverse momentum $E$ and a medium 
modified fragmentation function for the final hadron [$\tilde{D}^h(p_T/E)$], 
The modification of the partonic jet is encoded in the calculation of the 
medium modified fragmentation function. The four schemes of energy loss are 
in principle a different set of approximation schemes to estimate this quantity from 
perturbative QCD calculations.  

The reaction operator approach in opacity, often referred to as the Gyulassy-Levai-Vitev (GLV) scheme~\cite{GLV}, 
assumes the medium to be composed of heavy, almost static, 
colour scattering centers (with Debye screened Yukawa potentials) 
which are well separated in the sense that the mean free path of a jet 
$\lambda \gg 1/\mu$, the colour screening length of the medium. 
The opacity of the medium $\bar{n}$ quantifies the number of scattering centers seen  by a 
jet as it passes through the medium, \tie, $\bar{n} = L/\lambda$, where $L$ is the 
thickness of the medium. 
At leading order in opacity, a hard jet, produced locally in such a plasma with a large forward 
energy $E \gg \mu$, scatters off 
one such potential and in the process radiates a soft gluon. 
Multiple such interactions in a Poisson approximation are considered to calculate the 
probability for the jet to lose a finite amount of its energy. 

The path integral in opacity approach, referred to as the 
Armesto-Salgado-Wiedemann (ASW) approach~\cite{ASW}, also assumes a model for 
the medium as an assembly of Debye screened heavy scattering centers.
A hard, almost on shell,  parton traversing such a medium will 
 engender multiple transverse scatterings of  order $\mu \ll E$. It will in the 
process split into an outgoing parton and a radiated gluon which will also scatter 
multiply in the medium. 
The propagation of the incoming (outgoing) partons as well as that of the radiated gluon 
in this background colour field may be expressed in terms of  effective Green's functions, which 
are obtained in terms of path integrals over the field. Also similar to the GLV approach, a Poisson 
approximation is then used to obtain multiple emissions and a finite energy loss.

In the finite temperature field theory scheme referred to as the Arnold-Moore-Yaffe (AMY) approach~\cite{AMY}, 
the energy loss of hard jets is considered in an 
extended medium in equilibrium at asymptotically high temperature $T \ra \infty$ (and as a result $g \ra \infty$). 
In this limit, one uses the effective theory of hard-thermal-loops (HTL) to describe the collective properties of the 
medium.
A hard on-shell parton undergoes soft scatterings with momentum transfers $\sim gT$ off other hard 
partons in the medium. Such soft scatterings induce collinear radiation from the parton, with 
a transverse momentum of the order of $g T$. 
Multiple scatterings of the incoming (outgoing) parton 
and the radiated gluon need to be considered to get the leading order gluon radiation rate.
This is obtained from the imaginary parts of infinite order ladder diagrams. 
These rates are then used to evolve an initial distribution of hard partons through the medium 
in terms of a Fokker-Plank equation. 

In the higher-twist scheme~\cite{HT}, one directly computes the modification to the fragmentation functions  
due to multiple scattering in the medium by identifying and re-summing a class of higher twist 
contributions which are enhanced by the length of the medium. The initial hard jet is assumed to 
be considerably virtual, with $Q \gg \mu$. The propagation and collinear gluon emissions from such a 
parton are influenced by the multiple scattering in the medium. One assumes that, on exiting the medium, 
the hard parton has a small, yet perturbative scale $q^2$. One evolves this scale back up to the hard 
scale of the original produced parton, $Q^2$, by including the effect of multiple emissions in the medium. 
The multiple scatterings introduce a 
destructive interference for radiation at very forward angles and as such modify the evolution of the 
fragmentation functions in the medium. 

In any scheme, the magnitude of the modification is controlled by a single space-time dependent 
parameter which may be 
related to the well known transport coefficient $\hat{q}$. This is defined as the mean transverse momentum squared per 
unit length, transferred by the medium to the hard jet.  In actual computations, a model of the space-time dependence 
is invoked, 
and a maximum for $\hat{q}$ is set to best fit with experimental results. 
Values of the maximum of 
$\hat{q}$, in the vicinity of the center of a central collision, at a time of $\sim 1$fm/c, range from 
1 GeV$^2$/fm up to 20 GeV$^2$/fm~\cite{Majumder:2007iu}, depending on the scheme, as well as, the 
model of the medium used.

\section{Jet Correlations}

\begin{figure}[htbp]
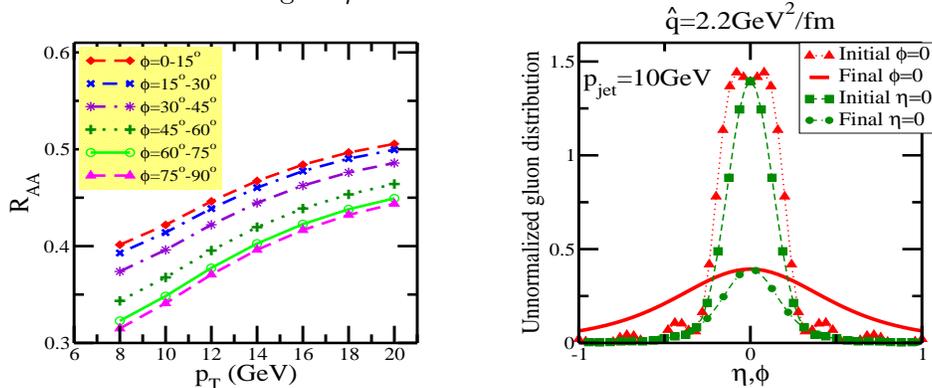

\mbox{}
\hspace{0.5cm}
\resizebox{1.95in}{1.75in}{\includegraphics[0.5in,0.5in][4.5in,4.6in]{r_aa_vs_reac_3.eps}}
\hspace{1.5cm}
\resizebox{1.95in}{1.75in}{\includegraphics[0.5in,0.5in][4.5in,4.6in]{cross_phi_and_eta_10.eps}}
\caption{Left panel: $R_{AA}$ as a function of $p_T$ for different ranges of $\phi$ the angle to 
the reaction plane in $Au$-$Au$ collisions 
at $20-30\%$ centrality. 
Right Panel: The pseudo-rapidity ($\eta$, at $\phi = 0$) and azimuthal angle ($\phi$, at $\eta = 0$) distribution of 
4 GeV gluons radiated from a 10 GeV trigger quark immediately after formation (Initial) and after propagating through $3$ fm of 
turbulent plasma (Final).}
\label{fig1}
\end{figure}

While the suppression of single inclusive hadrons may be used to determine the maximum value of $\hat{q}$, 
corrrelations between the leading hadron and the medium  may be 
used to test the space-time profile which is used as the ansatz~\cite{Majumder:2006we}. Once considers the 
nuclear modification factor $R_{AA}$ as a function of the angle with the reaction plane. The results of 
such an analysis, from the higher twist approach, for a medium profile taken from a full $3-D$ hydrodynamics 
simulation at $20-30$ \% centrality are presented in the left panel of Fig.~\ref{fig1}. The plots clearly demonstrate 
that the  modification is maximal when the jet propagates through the thickest part of the medium in a direction 
perpendicular to the reaction plane. The spread between the lines is directly dependent on the particular space-time ansatz 
chosen for the evolution of the produced matter. 

To determine the momentum structure of the plasma, one generalizes  $\hat{q}$ from a 
scalar to a tensor~\cite{Majumder:2006wi}, where the scalar 
$\hat{q} = \kd^{ij} \hat{q}_{ij}$. 
An example where such a situation may arise is in the 
presence of  large  turbulent colour fields, which may be generated in the early 
plasma due to anisotropic parton distributions~\cite{Asakawa:2006tc}. These 
large fields, transverse to the beam, tend to deflect radiated gluons 
from a transversely traveling jet, preferentially, in the longitudinal directions. 
While such effects influence the solid angle distributions of the radiated gluons 
around the originating parton, they do not have a considerable effect on the 
total energy lost. 
Such phenomena may yield an explanation for the ridge like structure seen in the 
near side correlations~\cite{Majumder:2006wi}. In the right panel of Fig.~\ref{fig1}, 
results from a quantitative estimate are presented where the trigger quark (with $E=10$ GeV) 
radiates a 4 GeV gluon, 
which is then subjected to such transverse fields (over a distance of 3 fm). Using a time dependent $\hat{q}$, with 
an initial value consistent with fits to experiment, one obtains a noticeable ridge like 
structure, \tie, a considerable broadening in $\eta$ but not in $\phi$.  

\end{document}